**Use of research-based instructional strategies: how to avoid faculty quitting**

Carl Wieman[*], Louis Deslauriers,
Department of Physics

Brett Gilley
Department of Earth, Ocean, and Atmospheric Sciences
University of British Columbia, Vancouver, B.C., Canada, V6T2L2

Abstract
We have examined the teaching practices of faculty members who adopted research-based instructional strategies as part of the Carl Wieman Science Education Initiative (CWSEI) at the University of British Columbia. Of the 70 that adopted such strategies with the support of the CWSEI program, only one subsequently stopped using these strategies. This is a tiny fraction of the 33% stopping rate for physics faculty in general [Henderson, Dancy, and Niewiadomska-Bugaj, PRST-PER, 8, 020104 (2012)]. Nearly all of these UBC faculty members who had an opportunity to subsequently use RBIS in other courses (without CWSEI support) did so. We offer possible explanations for the difference in quitting rates. The direct support of the faculty member by a trained science education specialist in the discipline during the initial implementation of the new strategies may be the most important factor.

PACS numbers: 01.40.Fk, 01.40.gb

---

[*] Corresponding author, gilbertwieman@gmail.com

Henderson, Dancy, and Niewiadomska-Bugaj surveyed 722 physics faculty as to their knowledge and use of research-based instructional strategies[1] (RBIS). In this important and informative paper, they were particularly concerned with what factors were important in the decision to use and to stop using such strategies. They found that a third of the faculty that reported using research-based instructional strategies reported that they subsequently quit using them. That inspired us to collect data on the number of faculty adopters and quitters of RBIS connected with the Carl Wieman Science Education Initiative (CWSEI) at the University of British Columbia. We found that the fraction that adopted and then stopped using RBIS was very low. In this brief note, we report the results and offer some possible reasons for the differences in the fraction of quitters in our sample compared the sample in Ref. [1].

## I. BACKGROUND

As described in Ref. [2], the CWSEI is a program designed to achieve widespread improvement of teaching and learning of science at the University of British Columbia. It is a competitive grant program that funds most, but not all, of the UBC science departments to carry out departmental-wide transformation of undergraduate courses. The transformed courses incorporate improved learning goals, research-based instructional methods (or "strategies" to use the terminology of Ref. [1] , and improved assessments of learning. Grants depend on the size of the department and the scope and quality of the proposal, but have been about $1.7 M over six years for large departments, such as Physics and Astronomy (PHAS) and Earth, Ocean, and Atmospheric Sciences (EOAS). EOAS has had a major grant for 5.5 years, PHAS for about 5, and Life Sciences for about 2.5 years. Computer Sciences has been funded for 5 years but has ramped up more slowly. All the departments have chosen to use most of the money to hire Science Education Specialists (SES). These are people who are highly knowledgeable in the respective disciplines, typically Ph.D.s, who are trained in instructional methods and discipline-based science education research by the CWSEI central staff and other SESs. The SESs then work sequentially with individual faculty, or small groups of faculty, to transform courses taught by those faculty, and correspondingly, the instructional strategies used by those faculty. The duties and training of the CWSEI SESs are discussed in more detail in Ref. [2].

The SESs have two primary modes of working with faculty. In the first, they have an intensive ongoing collaboration with one or more faculty to transform a particular course. The collaboration typically lasts for two iterations of the transformed course. The SESs help develop course materials such as learning goals, clicker questions, worksheets for in-class group problem solving, and homework problems. Then they work closely with the faculty members to implement the use of RBIS in class. They attend many of the classes, interview students, sometimes helping directly with implementation, and always offering feedback to the faculty member. In the second mode, the SESs serve as consultants rather than ongoing collaborators. Faculty will come to them wanting to get suggestions for RBIS they can use to solve specific problems, or to get guidance on how to implement a particular RBIS in a particular course. The SESs will offer advice and references but not ongoing assistance.

The instructional strategies introduced vary somewhat with the course and desires of the faculty, but typically include most, and often all of: targeted pre-class reading assignments and online quizzes (sort of "JITT-lite"), Peer Instruction with clickers, and in-class collaborative problem solving with worksheets or other formats. Most of the PHAS courses use PHET simulations, and many have introduced various forms of interactive lecture demonstrations and pre-post testing using standard physics instruments (FCI, BEMA, CLASS). Other departments have often developed their own diagnostic tests and/or pre-post tests similar to the CLASS and concept inventories. Thus, there is a large degree of overlap between the RBIS listed on the survey of Ref. [1] and those introduced in CWSEI transformed courses. There have also been new or improved TA training programs introduced in the departments and a more coordinated role of TAs in the courses.

## II. DATA

To determine the number of adopters and quitters of RBIS in the CWSEI affiliated departments at UBC, we first collected from all of the SESs the number of faculty with whom they had worked to help incorporate RBIS into their instruction.  Then we eliminated from those lists all the faculty members that had not been "on their own" for at least one academic year after the end of the time when they had formally worked with the SES. This period was chosen to be a reasonable minimum interval during which they could decide to quit using RBIS. This criterion eliminated a substantial number of faculty members who had adopted RBIS with CWSEI support, particularly in the life sciences where the program is relatively young.  We also eliminated any faculty members from the sample that were not in a position to use RBIS subsequent to their work with an SES, because they had retired, were on leave,  or had not taught any undergraduate courses.  We then checked if the members of the remaining sample of faculty had continued to use RBIS in largely the same manner as they had adopted with SES support.  That information was provided by self-reports of the faculty in a few cases, and from reports by the SESs in most cases, based on their recent discussions with the faculty members about their teaching strategies, and/or observations of the courses.  The SESs tend to keep track of what teaching strategies are being used in courses they helped transform, and so in many cases this information was immediately available.  Finally, there were several faculty members in computer science who were known to have adopted RBIS more than a year ago, and their current instructional strategies were also known from discussions with the current SESs, but, because of a turnover in SES personnel, it was unclear as to the extent of CWSEI assistance in their adoption of RBIS.  We included those faculty members in our sample. We have no independent confirmation of the accuracy of these reports, although we did set the criteria for the SESs that the RBIS being used had to be "largely the same" as what was introduced.  In most cases, the faculty members reported that RBIS was being used just as it had been at the completion of the SES involvement.  That involvement would typically have constituted one term of introduction followed by one term of refinement.

There were six departments that had faculty that met all the criteria listed above, PHAS, EOAS, Computer Science, and the three biology departments involved in jointly teaching the undergraduate life sciences program.  The data from these departments are shown in Table I.  The variation in the number of faculty from the respective departments that met the criteria is a reflection of the differences in the duration of funding and rate of ramp-up of the respective CWSEI activities.

| Department | # RBIS adopters | # RBIS quitters | # used in 2nd course/ # with opportunity |
|---|---|---|---|
| EOAS | 30 | 1 | 23/25 |
| PHAS | 25 | 0 | 23/25 |
| Computer Science | 9 | 0 | no data |
| Life Sciences | 6 | 0 | no data |

**Table I**. Number of faculty that adopted RBIS, and number that adopted RBIS but subsequently quit using them, per the criteria defined in the text, and number of faculty that had the opportunity to use RBIS in additional course(s) and the number of those that did so, for EOAS and PHAS.

In addition to the categories of faculty discussed above, there were three faculty members that we categorize as "encouraged, but never started." These are faculty who were assigned by the department to work with an SES to transform a course and introduce RBIS, but from the beginning the faculty member indicated great reluctance and discomfort with RBIS, refusing to use it at all, or trying it a little, and then stopping before the term was completed. This small number includes at least one faculty member whose fluency in English was limited and so was far more comfortable giving a pre-prepared lecture that they could write out ahead of time. While one might characterize these as "quitters", we do not think they are comparable to the quitters in Ref. [1], as these faculty expressed considerable reluctance from the beginning, and the only reason they had any participation was due to departmental pressure. This is a very unusual situation for a university faculty member, and we do not believe this would be representative of the quitters in Ref. [1].

There were some other faculty members in every department who did not believe in RBIS and simply refused to use them. The fraction of the faculty with those beliefs has noticeably decreased over time in every department that has a CWSEI supported initiative. In EOAS, where there has been the most widespread involvement, with about ¾ of the faculty having worked with an SES, it is now nearly zero, and in PHAS it has dropped markedly with time, although it still remains a significant fraction. We have the impression it may be higher in the other departments but do not have data on this.

Our sample includes faculty members from multiple disciplines at a large public research university in Canada, in contrast to the sample of Ref. [1], which was limited to physics faculty at a range of institutions in the US. We do not believe that these differences are relevant to the reasons why faculty quit using RBIS. In working extensively with the faculty from the departments listed, we have seen little if any difference between the PHAS and non-PHAS faculty in terms of their attitudes about teaching in general and specifically the adoption of RBIS. There was a slightly higher initial level of knowledge about RBIS in PHAS compared to other departments, but the rapid overall increase in knowledge due to the CWSEI activities quickly dwarfed that initial difference. With regard to the country difference, one of us (CEW) has worked extensively at and with US universities, and at the level of the departments and individual faculty members, UBC looks very similar to a comparable large US research university. This is not surprising as a large fraction of the UBC science faculty have attended and/or have been faculty members at US institutions, and there is a large amount of cross-border travel, interaction, and collaborations in the sciences. The pre-CWSEI level of knowledge and use of RBIS in PHAS was somewhat lower than the average reported in Ref. [1], with only about 10% using any RBIS, but was not that different from many physics departments in the US. We also have anecdotal reports of several faculty having tried RBIS without CWSEI support, and then stopped using them, not unlike those cases discussed in Ref. [1].

### III. DISCUSSION

The fraction of this sample of UBC faculty who used RBIS and then quit is 1 out of 70, or less than 1/20[th] of what Henderson, Dancy, and Niewiadomska-Bugaj found[1]. It is worthwhile to compare the contexts in which the two sample populations are adopting RBIS to understand what factors might explain this dramatic difference. This provides insight as to what kinds of support would best help faculty adopt and continue to use RBIS. The two most significant factors appear to be:

- Help from SESs. The SESs are trained as to best practices and potential challenges with the implementation of these techniques, and hence, how to avoid problems in the first place. They also help tailor the use of RBIS to the specific course context, and they are present during the implementation to provide immediate help and advice in overcoming problems and improving execution.

- A supportive departmental environment in which the department has made a commitment to transform how courses are taught, and many faculty are using RBIS and are regularly exchanging ideas and discussing the successes and challenges, including frequent input to those discussions from the SESs.

We believe that, while both factors are significant, the first is the most important. This is consistent with Ref. [3]. Having a knowledgeable person who can minimize the initial challenges of implementation and ensure that RBIS are successful and well received by students when first implemented is an enormous step towards encouraging faculty to embrace the use of RBIS. As well as helping deal with the general challenges and potential pitfalls of implementation, the SESs also provide expert help to the faculty members on how to adapt specific RBIS to their particular situation, needs, and course material. When RBIS is implemented successfully, the resulting enhanced student engagement and learning are then apparent, making the experience rewarding for the faculty. This is very frequently reflected in the feedback the faculty provide to the SESs on the experience. If further research shows that this particular support is in fact the dominant factor in successful adoption and continued use of RBIS, it is likely that it can be provided in less expensive ways than the CWSEI model.

The SESs share amongst themselves the wisdom gained about the challenges and techniques for successful implementation of RBIS, and the most effective sorts of feedback to provide to faculty members as they are first implementing RBIS. Much of the former is contained in the guides that are posted on the CWSEI website under "Resources: Instructor Guidance". However, such a guide is very different from having someone in the classroom that can help at a critical moment.

Another type of support that the SESs provide that might be significant is the theoretical underpinnings as to why these RBIS are effective for learning. The SESs are trained in the research on learning and the principles behind the various RBIS, and this is communicated informally to the faculty.

The influence of more general departmental factors is less clear. On the one hand, there is clearly departmental encouragement and support for adopting RBIS. The departments that receive CWSEI funding have Heads who have explicitly supported these efforts (that is a requirement for support), and the departments have offered a variety of incentives (course buyouts, extra TAs, etc.) to faculty to participate in the course transformation efforts. Also, as the efforts have progressed, there is greater discussion, often actively facilitated by the SESs, and interest within the department in RBIS, encouraging more faculty to try them. Undergraduate students begin to express their expectations of RBIS in their classes. TA training programs have also been instituted that focus heavily on RBIS. Finally, in some large courses with multiple instructors, clear expectations have been established that any faculty member assigned to teach this course will use particular RBIS. All of these must encourage the continued use of RBIS.

On the other hand, this "support" often becomes "pressure". To receive CWSEI funding, departments have to make specific commitments, and the CWSEI leadership and the Dean regularly review their progress towards meeting those commitments, and to some extent this "attention" is passed through to the individual faculty. So, relative to the sample population in Ref. [1], this means that many faculty are trying RBIS who would not have, if they were in a typical department. There are a numerous examples of faculty members who are working with SESs and adopting RBIS in response to direct pressure from the departmental administration. These faculty likely have bought into the concept less than those who spontaneously adopt RBIS with little departmental encouragement, such as most of the sample in Ref. [1]. One might expect that faculty members who are adopting RBIS in response to such departmental pressure would be more likely to quit later. The overall results indicate however, that even those faculty who are pressured into trying RBIS will embrace it and continue to use it, if they have the support of an SES in the process.

We have also briefly explored two other aspects of RBIS use among the faculty in this sample, in response to suggestions by Melissa Dancy. The first is how well is the fidelity of the RBIS being maintained over time? This question was posed to the SESs, along with a specific set of criteria to use in determining fidelity, such as, if Peer Instruction was being used, was the student-student discussion preserved? With three caveats, the fidelity appears to be maintained quite well; with little change from when faculty were using RBIS with the help of the SESs. Some faculty members have dropped one particular RBIS of several they were using, but they continue to use the others with fidelity, while some faculty have added RBISs that they had not previously used.

The first caveat is that the data is limited. Although the SESs talk informally with many faculty on a regular basis and observe classes, we do not have solid data on the fidelity of use by all the faculty members. The second caveat is that, in a few cases, faculty have felt compelled to get through a departmentally set syllabi and gave up many RBIS near the end of the term as they rushed to get through the material. They complained about this situation and went back to using RBIS with full fidelity at the start of the next term. The third caveat is that the duration of the RBIS use after the end of direct SES support is necessarily limited by the age of the CWSEI. In most cases, it is three years or less. Although most of the faculty members in this sample have expressed enthusiasm about the RBIS as they are currently using them, it is possible that fidelity may decay over time.

The second question Melissa raised is, how often are faculty that started using RBIS with CWSEI support introducing RBIS into new courses when there is no CWSEI support for their efforts? We examined this for the faculty adopters of RBIS in the departments of EOAS and PHAS. For approximately 2/3 of the cases, the faculty member or students had already told one of us that RBIS had been introduced in a new course. We surveyed the remaining faculty. Twenty-three of the 25 RBIS users on the PHAS faculty were found to be using RBIS in additional courses. Four of the 29 non-quitting EOAS faculty have not taught a different undergraduate course. Of the remaining 25, 23 have implemented RBIS in additional courses. It should be noted that many of these are self-reports, and the faculty know what we would like to hear. Nevertheless, it is a large number, larger than we had expected. We have also heard of a number of examples of faculty members applying RBIS to new courses in computer science and life sciences, but do not have meaningful data on the extent.

In conclusion, we have found that only one of the 70 UBC faculty members in science that adopted RBIS with the support of the CWSEI and might have stopped using RBIS, did so. This is a dramatic contrast with the finding of Henderson, Dancy, and Niewiadomska-Bugaj[1] that one third of the physics faculty that adopted RBIS later stopped using them. In addition, over 90% of the UBC faculty adopters of RBIS in the departments of EOAS and PHAS later starting using RBIS in other courses, when they had the opportunity, with minimal or no CWSEI support. Our data provides a considerably more optimistic outlook for the adoption of RBIS. It implies that with good support, nearly all faculty members can successfully adopt and happily continue using RBIS. While this demonstrates what can be achieved, an important area of future research will be to determine precisely what support is the most important of everything that is provided by the CWSEI, and what is the most cost effective way to provide that support.

**ACKNOWLEDGEMENTS**

We are pleased to acknowledge valuable discussions with Charles Henderson and Melissa Dancy that both inspired us to write this paper and contributed to the content, and the assistance of the CWSEI SESs who provided data, Costanza Piccolo, Warren Code, Joseph Lo, Ed Knorr, Josh Caulkins, Mandy Banet, Bridgette Clarkson, Laura Weir, Lisa Macdonald, Cynthia Heiner, Jim Carolan, and Peter Newbury. This work was supported by the University of British Columbia through the CWSEI.